# A Performance Analysis of HICCUPS – a Steganographic System for WLAN

Krzysztof Szczypiorski
*Warsaw University of Technology, Institute of Telecommunications*
*ul. Nowowiejska 15/19, 00-665 Warsaw, Poland*
E-mail: `ksz@tele.pw.edu.pl`

**Abstract**

The paper presents an analysis of performance features of the HICCUPS (*HIdden Communication system for CorrUPted networkS*) including the efficiency and the cost of the system in WLANs (*Wireless Local Area Networks*). The analysis relies on the original CSMA/CA (*Carrier Sense Multiple Access with Collision Avoidance*) 802.11 Markov chain-based model.

## 1. Introduction

The HICCUPS (*HIdden Communication system for CorrUPted networkS*), introduced by the author in [6], is a steganographic system for WLANs (*Wireless Local Area Networks*). The main innovation of the system is usage of frames with intentionally wrong checksums to establish covert communication. The HICCUPS was recognized [1] as the first steganographic system for WLAN.

The analysis presented in this paper focuses on some performance features of the HICCUPS, including the efficiency and the cost of the system usage in WLAN. For the purpose of this analysis the Markov chain-based model was used which is dedicated for 802.11 CSMA/CA (*Carrier Sense Multiple Access with Collision Avoidance*; [7],[3],[4],[5],[2]). The cost of system usage ($\kappa$) is defined as a decline of WLAN throughput that results from the HICCUPS operating in the corrupted frame mode [6]. The efficiency of the system ($\varepsilon$) is defined as a throughput of the system in the corrupted frame mode.

The evaluation was performed for the saturated condition i.e. when all stations involved in communications have no empty queues. Saturation throughout ($S$) is an efficiency measure of maximum load in saturated conditions.

This work is based on the author's PhD thesis [7].

## 2. The Analysis of Saturation Throughput for the Corrupted Frame Mode – $S_H$

### 2.1 Calculation of $S_H$

First we evaluate the saturation throughput for the HICCUPS in the corrupted frame mode ($S_H$). The analysis is similar to effort done for the 802.11 CSMA/CA networks in [7],[3],[4],[5],[2].

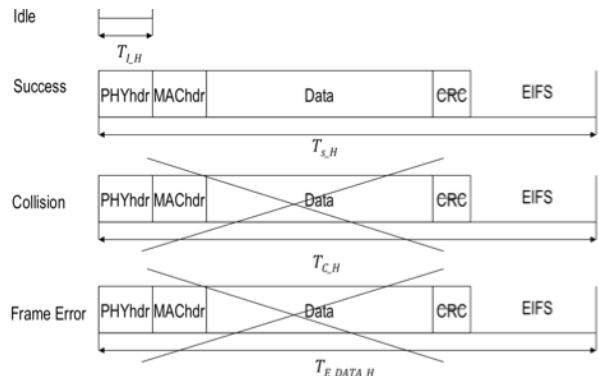

**Figure 1. States of the channel**

Figure 1 illustrates four states of the channel that could occur during the corrupted frame mode. In this mode all 802.11 frames have incorrect value of CRC-32 code deliberately set in the FCS field (*Frame Checksum Control*). Thus, there are no positive acknowledgments through ACK (*ACKnowledgment*) frames, and therefore „ACK error" state is ommited ([7],[3],[4],[5],[2]). The „success" of the transmission in the HICCUPS, not defined in the same way as for the 802.11 network, means that during transmission there were no collisions and no data errors. The mechanism of frame integrity for the HICCUPS is separate from 802.11 FCS.

The duration of four states are as following (Fig. 1):

$T_{I\_H}$ – idle slot,
$T_{S\_H}$ – successful transmission,
$T_{C\_H}$ – transmission with collision,

$T_{E\_DATA\_H}$ – unsuccessful transmission with data frame error.

So we have:

$$\begin{cases} T_{I\_H} = \sigma \\ T_{S\_H} = T_{PHYhdr} + T_{DATA} + \delta + T_{EIFS} \\ T_{C\_H} = T_{S\_H} \\ T_{E\_DATA\_H} = T_{S\_H} \end{cases} \quad (1)$$

Probabilities corresponding to states of the channel are denoted as follows:
$P_{I\_H}$ – probability of idle slot,
$P_{S\_H}$ – probability of successful transmission,
$P_{C\_H}$ – probability of collision,
$P_{E\_DATA\_H}$ – probability of unsuccessful transmission due to data frame error.

Let $\tau_H$ be a probability of frame transmission in the corrupted frame mode, $p_{e\_data}$ a probability of data frame error (see the formula (22) in [3]). These are related to channel state probabilities as follows (see the equation (12) in [3]):

$$\begin{cases} P_{I\_H} = (1-\tau_H)^n \\ P_{S\_H} = n\tau_H(1-\tau_H)^{n-1}(1-p_{e\_data}) \\ P_{C\_H} = 1-(1-\tau_H)^n - n\tau(1-\tau_H)^{n-1} \\ P_{E\_DATA\_H} = n\tau_H(1-\tau_H)^{n-1}p_{e\_data} \end{cases} \quad (2)$$

We use the same assumptions as stated in chapter 2.1 of [3], so we could express $S_H$ (similar to the formula (6) in [3]):

$$S_H = \frac{P_{S\_H}L_{pld}}{T_{I\_H}P_{I\_H} + T_{S\_H}P_{S\_H} + T_{C\_H}P_{C\_H} + T_{E\_DATA\_H}P_{E\_DATA\_H}}, \quad (3)$$

where $L_{pld}$ is a length of data in frame with FCS field, expressed in bps. $S_H$ could be normalized to $R$ - the rate of the 802.11 network (see formula (7) in [3]):

$$\bar{S}_H = \frac{S_H}{R} \quad (4)$$

## 2.1 Probability of Frame Transmission in the Corrupted Frame Mode - $\tau_H$

Based on the model presented and evaluated in [7],[3],[4],[5],[2] let us consider a model of the 802.11 CSMA/CA backoff procedure in corrupted frame mode. From a WLAN perspective of the HICCUPS, communication always fails, because of absence of proper checksums. Hence transmission of steganograms is performed in every step of the backoff procedure, so we could describe the HICCUPS behaviour with the Markov chain-based model as presented in [7],[3],[4],[5],[2]) with probability of the failure $p_f$=1 (means "always failure").

The state of the two-dimensional process $(s(t), b(t))$ will be denoted as $(i,k)$ ([7],[3],[4],[5],[2]), $b_{i,k}$ is a probability of this state. The one-step conditional state transition probabilities will be denoted by $P = (\bullet,\bullet|\bullet,\bullet)$.

Non-full transition probabilities are determined as follows:

$$\begin{cases} P(i,k|i,k+1) = 1-p_{coll}, & 0 \le i \le m, 0 \le k \le W_i - 2 \\ P(i,k|i,k) = p_{coll}, & 0 \le i \le m, 1 \le k \le W_i - 1 \\ P(i,k|i-1,0) = 1/W_i, & 0 \le i \le m, 0 \le k \le W_i - 1 \\ P(0,k|m,0) = 1/W_0, & 0 \le k \le W_0 - 1 \end{cases} \quad (5)$$

where $p_{coll}$ is a probability of collision, $W_0$ is an initial size of th contention window and $m'$ is a maximum number by which the contention window may be doubled; $m'$ may be both greater and smaller than $m$ and also equal to $m$. $W_i$ is the maximum value of a backoff timer at the $i$ backoff stage:

$$W_i = \begin{cases} 2^i W_0, & i \le m' \\ 2^{m'}W_0 = W_m, & i > m' \end{cases} \quad (6)$$

With transition probabilities as above (5) and justifications as in ([7],[3],[2]), Markov chain transitions is presented in Fig. 2. Let us notice that differences between this diagram and the 802.11 CSMA/CA diagram ([7],[3],[4],[5],[2]) are lack of returns to states $(0,k)$ for $0 \le k \le W_0$-1 and $(i,0)$ for $0 \le i \le m$-1 – this is a graphical interpretation of "always failure" from the perspective of WLAN.

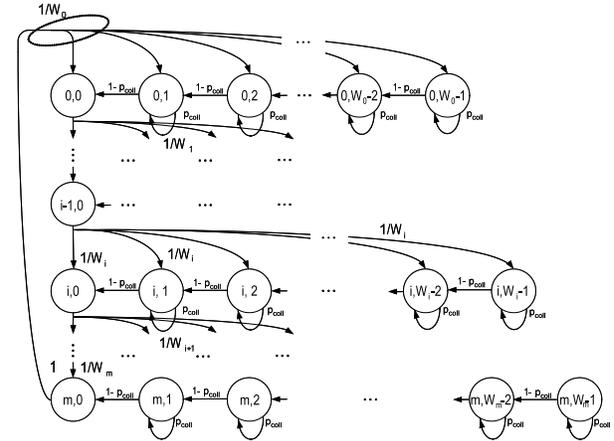

**Figure 2. Markov chain transitions**

For $0 \le i \le m$ we have:

$$b_{i,k} = \begin{cases} \dfrac{W_i - k}{W_i(1-p_{coll})}b_{0,0}, & 0 < k \le W_i - 1 \\ b_{0,0}, & k = 0 \end{cases} \quad (7)$$

Because

$$\sum_{i=0}^{m} b_{i,0} = b_{0,0}(m+1) \quad (8)$$

and (7) we get:

$$1 = \sum_{i=0}^{m}\sum_{k=1}^{W_i-1} b_{i,k} + \sum_{i=0}^{m} b_{i,0} = \frac{b_{0,0}}{1-p_{coll}} \sum_{i=0}^{m} \frac{W_i-1}{2} + b_{0,0}(m+1) \cdot \quad (9)$$

and

$$b_{0,0}^{-1} = \begin{cases} \frac{W_0(2^{m+1}-1)-(m+1)}{2(1-p_{coll})} + (m+1), & m \le m' \\ \frac{W_0(2^{m'+1}-1)-(m+1)+(m-m')W_0 2^{m'}}{2(1-p_{coll})} + (m+1), & m > m' \end{cases} \quad (10)$$

Having $b_{0,0}$ we may calculate (similar to [7],[3],[4],[5],[2]) probability of frame transmission in the corrupted frame mode:

$$\tau_H = \sum_{i=0}^{m} b_{i,0} =$$
$$= \begin{cases} \left(\frac{W_0(2^{m+1}-1)-(m+1)}{2(1-p_{coll})} + (m+1)\right)^{-1}(m+1), & m \le m' \\ \left(\frac{W_0(2^{m'+1}-1)-(m+1)+(m-m')W_0 2^{m'}}{2(1-p_{coll})} + (m+1)\right)^{-1}(m+1), & m > m' \end{cases} \quad (11)$$

Probability $p_{coll}$, similar to the formula (25) in [3], is:

$$p_{coll} = 1 - (1-\tau_H)^{n-1} \cdot \quad (12)$$

Equations (10) and (11) form a system with two unknown variables $\tau_H$ i $p_{coll}$ which may be solved numerically.

## 3. The Cost – $\kappa$

According to the definition of the cost ($\kappa$), introduced in the first part of this paper, the cost is the difference between $S$, for frame error rate without the HICCUPS, and $S$, with frame error rate as a result of the HICCUPS in the corrupted frame mode. In other words $\kappa$ is a decline of WLAN throughput grabbed by HICCUPS hidden channels.

Let us assume that the HICCUPS increases frame error rate by the constant value $\Delta FER$ (Fig. 3) and frame error rate of the networks without the HICCUPS equals $FER'$. We could notice that $0 \le \Delta FER \le 1-FER'$. So we could express the cost as:

$$\kappa = S(FER') - S(FER'+\Delta FER) \quad (13)$$

and normalized to $R$:

$$\bar{\kappa} = \frac{\kappa}{R} \cdot \quad (14)$$

The curves of the cost are based on $S(FER)$ and they look almost linear [7], so for small values of $\Delta FER$ we could use the following approximation formula (Fig. 4):

$$\kappa \approx \frac{\Delta FER}{1-FER'} RPN_{WLAN}(FER') \cdot \quad (15)$$

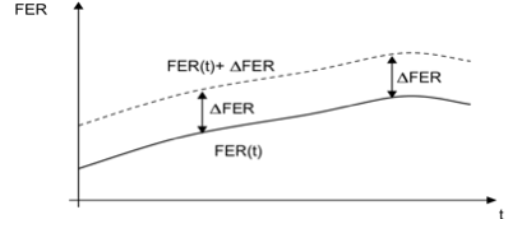

**Figure 3. Interpretation of $\Delta FER$**

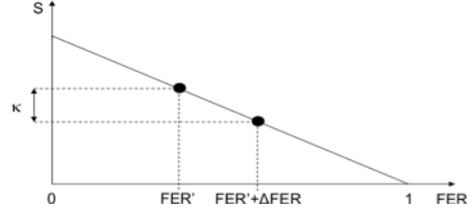

**Figure 4. Graphical presentation of the cost ($\kappa$)**

In tables 1 and 2 the values of the cost $\kappa$ for $n=5$ and $n=10$ are presented for IEEE 802.11g (ERP-OFDM) 54 Mbps - [4][5]. These results, for $L=1000$ bytes, come from (15), and were calculated for $FER' \in \{0; 0.0769; 0.5507\}$ (that corresponds to three bit error rates: $BER \in \{0, 10^{-5}, 10^{-4}\}$). For these conditions five typical values of $\Delta FER$ were taken into account (0.01; 0.02; 0.03; 0.04; 0.05).

**Table 1. Normalized values of the cost $\kappa$ (in brackets expressed in Mbps) – $n=5$ and $L=1000$ bytes**

| $\Delta FER$ / $FER'$ | 0.01 | 0.02 | 0.03 | 0.04 | 0.05 |
|---|---|---|---|---|---|
| 0 | 0.0048 (0.26) | 0.0097 (0.52) | 0.0145 (0.78) | 0.0194 (1.05) | 0.0242 (1.31) |
| 0.0769 | 0.0049 (0.26) | 0.0097 (0.52) | 0.0146 (0.79) | 0.0194 (1.05) | 0.0243 (1.31) |
| 0.5507 | 0.0047 (0.25) | 0.0093 (0.50) | 0.0140 (0.75) | 0.0186 (1.01) | 0.0233 (1.26) |

**Table 2. Normalized values of the cost $\kappa$ (in brackets expressed in Mbps) – $n=10$ and $L=1000$ bytes**

| $\Delta FER$ / $FER'$ | 0.01 | 0.02 | 0.03 | 0.04 | 0.05 |
|---|---|---|---|---|---|
| 0 | 0.0046 (0.25) | 0.0092 (0.50) | 0.0138 (0.75) | 0.0184 (1.00) | 0.0230 (1.24) |
| 0.0769 | 0.0046 (0.25) | 0.0093 (0.50) | 0.0139 (0.75) | 0.0186 (1.00) | 0.0232 (1.25) |
| 0.5507 | 0.0047 (0.26) | 0.0095 (0.51) | 0.0142 (0.77) | 0.0190 (1.02) | 0.0237 (1.28) |

## 4. The Efficiency – $\varepsilon$

According to the definition of the efficiency ($\varepsilon$), as stated in the introduction, the efficiency is the $S_H$ in conditions that result from physical channel (especially its BER) and amount of frames used by the HICCUPS in the corrupted frame mode. These conditions enable different view on frame error rate from the HICCUPS perspective: the proper frames for the HICCUPS are corrupted for WLAN, and of course the good ones for WLAN in the meaning of the HICCUPS are wrong. So we will use $FER_H$ to emboss this difference, and define $\varepsilon$ as follows:

$$\varepsilon = S_H(FER_H) \quad (16)$$

$S_H$, evaluated in the first part of the paper, allows to calculate the upper boundary of HICCUPS throughput. In the normal use of the HICCUPS the corrupted frame mode occurs very rarely.

To estimate efficiency we might consider two scenarios. In the first scenario: all stations are in the corrupted frame mode only (the HICCUPS is always on): $S$ in the function of FER equals 0 (because $S(1)=0$), and $S_H$ in the function of FER equals $S_H(FER')$. Because $0 \leq \Delta FER \leq 1-FER'$, $\Delta FER=1-FER'$. In the second scenario: the HICCUPS is off ($\Delta FER=0$), only normal transmission is performed, so $S_H=0$ (because $S_H(1)=0$), $S$ equals $S(FER')$.

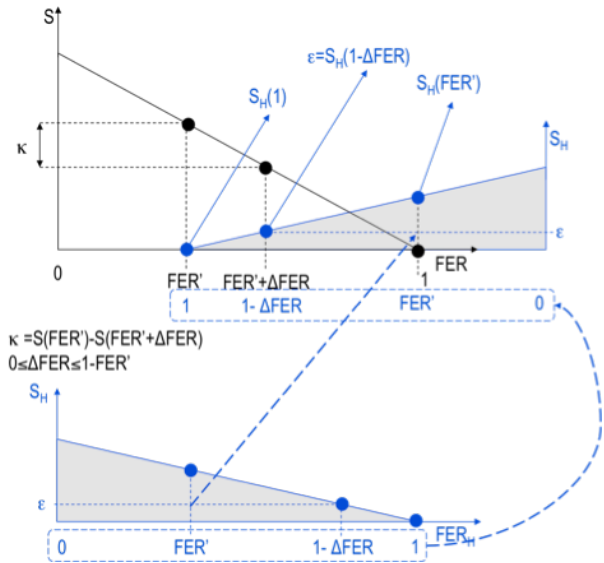

**Figure 5. Graphical interpretation of the efficiency $\varepsilon$**

On the base of the two scenarios presented above we could estimate the hypothetic point of the HICCUPS operation for ($FER'+\Delta FER$) as combination of the translation and the reflection (Fig. 5). The $S_H$ curve is reflected and then translated in FER domain to keep $S(1)=0$ and $S_H(FER')$ together as well as $S(FER')$ and $S_H(1)=0$. After this operations we could observe that $FER_H=1-\Delta FER$. Finally:

$$\varepsilon = S_H(1-\Delta FER) \quad (17)$$

and could be normalized to $R$:

$$\bar{\varepsilon} = \frac{\varepsilon}{R}. \quad (18)$$

Similarly to the analysis of the cost we consider an IEEE 802.11g (ERP-OFDM) 54 Mbps network with 1000 bytes frames, $n \in \{5,10\}$, and the same values of $\Delta$FER (0.01; 0.02; 0.03; 0.04; 0.05). The results are presented in table 3.

**Table 3. Normalized values of the efficiency $\varepsilon$ (in brackets expressed in Mbps) – $n \in \{5, 10\}$ and $L$=1000 bytes**

| $\Delta FER$ $n$ | 0.01 | 0.02 | 0.03 | 0.04 | 0.05 |
|---|---|---|---|---|---|
| 5 | 0.0042 (0.23) | 0.0085 (0.46) | 0.0127 (0.69) | 0.0169 (0.91) | 0.0212 (1.14) |
| 10 | 0.0047 (0.25) | 0.0094 (0.51) | 0.0141 (0.76) | 0.0188 (1.01) | 0.0235 (1.27) |

## 5. Conclusions

The analysis presented in this paper focuses on the performance features of the HICCUPS including the efficiency and the cost of system usage in WLAN. The analysis relies on the original Markov chain-based model. The cost depends on the frame error rate, and the efficiency depends only on $\Delta FER$. As an example for an IEEE 802.11g (ERP-OFDM) 54 Mbps network with 10 stations and $\Delta FER$=0.05, the efficiency $\varepsilon$ equals 1.27 Mbps and the cost $\kappa$ is 1.28 Mbps. The analysis proves that the HICCUPS is the efficient steganographic method with the reasonable cost.

Future work will focus on the simulation analysis of the HICCUPS to evaluate features of the systems in different scenarios and cover a versatile assessment of the HICCUPS security.

### References

[1] Krätzer, C., Dittmann, J., Lang, A., Kühne, T.: *WLAN Steganography: a First Practical Review*. In Proc. of: 8th ACM Multimedia and Security Workshop. Geneve (Switzerland). 26-27 September 2006

[2] Szczypiorski, K., Lubacz, J.: *Performance Analysis of IEEE 802.11 DCF Networks*. Journal of Zhejiang University - Science A, Zhejiang University Press, co-published with Springer-Verlag GmbH, Vol. 9, No. 10, October 2008. pp. 1309-1317

[3] Szczypiorski, K., Lubacz, J.: *Performance Evaluation of IEEE 802.11 DCF Networks*. In: Lorne Mason, Tadeusz Drwiega, and James Yan (Eds.) - Managing Traffic Performance in Converged Networks - Lecture Notes in Computer Science (LNCS) 4516, Springer-Verlag Berlin Heidelberg, Proc. of 20th International Teletraffic Congress - ITC-20, Ottawa, Canada, June 17-21, 2007. pp. 1084-1095

[4] Szczypiorski, K., Lubacz, J.: *Saturation Throughput Analysis of IEEE 802.11g (ERP-OFDM) Networks*. In: Robert Bestak, Boris Simak, and Ewa Kozlowska (Eds.) - Personal Wireless Communications - IFIP Vol. 245/2007, Springer Boston, Proc. of 12th IFIP International Conference on Personal Wireless Communications - PWC'07, Prague, Czech Republic, September 12-14, 2007, pp. 196-205

[5] Szczypiorski, K., Lubacz, J.: *Saturation Throughput Analysis of IEEE 802.11g (ERP-OFDM) Networks*. Telecommunication Systems: Modelling, Analysis, Design and Management, ISSN: 1018-4864 (print version), ISSN: 1572-9451 (electronic version), Springer US, Journal no. 11235 Vol. 38, Numbers 1-2, June, 2008. pp. 45-52

[6] Szczypiorski, K.: *HICCUPS: Hidden Communication System for Coruppted Networks*. In Proc: The Tenth International Multi-Conference on Advanced Computer Systems ACS'2003. Międzyzdroje. 22-24 October 2004. pp. 31-40

[7] Szczypiorski, K.: *Steganography in Wireless Local Area Networks*. PhD thesis (in Polish), Warsaw, September 2006, Warsaw University of Technology